\documentclass[twocolumn,superscriptaddress,showpacs,preprintnumbers,amsmath,amssymb]{revtex4}
\usepackage{graphicx}% Include figure files
\usepackage{dcolumn}% Align table columns on decimal point
\usepackage{bm}% bold math

\begin{document}

\title{Viscosity and dilepton production of a chemically equilibrating quark-gluon plasma at finite baryon density}

\author{ Nana Guan }
\affiliation{Shanghai Institute of Applied Physics, Chinese
Academy of Sciences, Shanghai 201800, China}
\affiliation{Graduate
School of the Chinese Academy of Sciences, Beijing 100080, China}

\author{Zejun He\footnote{E-mail address: hezejun@sinap.ac.cn}}
\affiliation{Shanghai Institute of Applied Physics, Chinese
Academy of Sciences, Shanghai 201800, China}

\author{Jiali Long }
\affiliation{Shanghai Institute of Applied Physics, Chinese
Academy of Sciences, Shanghai 201800, China}

\author{Xiangzhou Cai }
\affiliation{Shanghai Institute of Applied Physics, Chinese
Academy of Sciences, Shanghai 201800, China}

\author{Yugang Ma }
\affiliation{Shanghai Institute of Applied Physics, Chinese
Academy of Sciences, Shanghai 201800, China}

\author{ Jianwei Li }
\affiliation{Shanghai Institute of Applied Physics, Chinese
Academy of Sciences, Shanghai 201800, China}
\affiliation{Graduate
School of the Chinese Academy of Sciences, Beijing 100080, China}

\author{Wenqing Shen}
\affiliation{Shanghai Institute of Applied Physics, Chinese
Academy of Sciences, Shanghai 201800, China}

\date{\today}

\begin{abstract}

By considering the effect of shear viscosity we have investigated
the evolution of a chemically equilibrating quark-gluon plasma at
finite baryon density. Based on the evolution of the system we
have performed a complete calculation for the dilepton production
from the following processes: $q\bar{q}{\rightarrow}l\bar{l}$,
$q\bar{q}{\rightarrow}gl\bar{l}$, Compton-like scattering
($qg{\rightarrow}ql\bar{l}$,
$\bar{q}g{\rightarrow}{\bar{q}}l\bar{l}$),
 gluon fusion
$g\bar{g}{\rightarrow}c\bar{c}$, annihilation
$q\bar{q}{\rightarrow}c\bar{c}$ as well as the multiple scattering
of quarks. We have found that quark-antiquark annihilation,
Compton-like scatterring, gluon fusion, and multiple scattering of
quarks give important contributions. Moreover, we have also found
that the dilepton yield is an increasing function of the initial
quark chemical potential, and the increase of the quark phase
lifetime because of the viscosity also obviously raises the
dilepton yield.

\end{abstract}

\pacs{12.38.Mh, 25.75.-q, 24.10.Nz\\}

\maketitle

\section{\label{sec:level1}INTRODUCTION}

The Relativistic Heavy-Ion Collider (RHIC) at the Brookhaven
National Laboratory and the Large Hadron Collider (LHC) being
built at CERN will provide the best opportunity to study the
formation and evolution of quark-gluon plasma (QGP). Dileptons have large mean free path due
to the small cross section for
electromagnetic interaction in the plasma, therefore can provide an ideal probe for the
detection and study of the plasma.

Many authors [1-3], considering that the created QGP in collisions
is a thermodynamic equilibrium system, have studied the dilepton
production. Recently, the photon and dilepton productions were
studied based on the evolution model of chemically equilibrating
QGP, established by Shuryak, Bir\'o and co-workers [4-6],
 especially, the studies of these productions in the plasma at finite baryon density were also performed [7,8].
 However, most previous
 works were done by assuming the partonic plasma to be ideal, i.e., without any viscous effect.
 In principle, the viscous effects in fluid hydrodynamics should not be neglected in a realistic scenario
 since the dimension
 of the plasma is comparable to the mean free path of the partons. The viscous coefficient in the
 frame-work
 of hydrodynamics is composed of bulk and shear viscosity, while the bulk viscosity vanishes for a quark-gluon
 plasma [9].
 In this work, we mainly discuss the effect of the shear viscosity, which have attracted many authors to investigate
 its influences on the formation and evolution of the QGP system. Authors of [9] have studied the viscous
 corrections to
 the hydrodynamic equations describing the evolution of the QGP at finite
 baryon density, and investigated the effect of viscosity on chemical equilibration of the system. They have found that
 due to the viscosity the lifetime of the plasma increases, the temperature evolution of the system becomes slow, and
 the chemical equilibration of the system becomes fast, therefore, the reaction rate will be heightened.
 However, we should point out that
 in previous work many authors have regarded the viscous coefficients as
 adjustable parameters [9-12]. Indeed, they should be directly obtained from the thermodynamic quantities of the
 system. On the other hand, the viscous coefficients derived by Danielewicz and Gyulassy [13]
 based on QCD phenomenology
 for a baryon free-plasma, especially  by Hou and Li [14] considering the Debye screening and damping rate of gluons for a baryon-rich plasma
 using finite-temperature QCD, are so large that the temperature of the plasma would be abnormally heightened.

In early calculations, one mainly considered the dilepton
production from the process $q\bar{q}{\rightarrow}l\bar{l}$. In
recent years, possible sources of dileptons, such as
$q\bar{q}{\rightarrow}l\bar{l}$ annihilation,
$qg{\rightarrow}ql\bar{l}$ Compton-like scattering and
$qg{\rightarrow}qgl\bar{l}$ fusion, were investigated [15,16]. In
addition, the contributions of gluon fusion
$gg{\rightarrow}c\bar{c}$, quark-antiquark annihilation
$q\bar{q}{\rightarrow}c\bar{c}$ and multiple scattering of quarks
to dileptons have also been studied [17].

In this work, starting from the shear viscous coefficient
 given by relativistic kinetic theory for a massless QGP under relaxation time approximation,
 we first estimate the mean free paths of partons in a chemically equilibrating QGP at finite baryon density,
 then combining with the parton energy densities, calculate the shear viscous coefficient of the QGP.
Subsequently, based on our evolution model including the
viscosity, we perform a complete calculation for the dilepton
production from processes:
 $q\bar{q}{\rightarrow}l\bar{l}$,  $q\bar{q}{\rightarrow}gl\bar{l}$, Compton-like ($qg{\rightarrow}ql\bar{l}$,
$\bar{q}g{\rightarrow}{\bar{q}}l\bar{l}$),
gluon fusion $g\bar{g}{\rightarrow}c\bar{c}$, annihilation $q\bar{q}{\rightarrow}c\bar{c}$ as well as multiple
scattering of quarks to predict the contributions of these reaction processes,
and reveal the effect of the finite baryon density and viscous phenomena on dilepton production.

The rest of the paper is organized as follows: Sec. II describes
the evolution of the dissipative QGP system. In Sec. III, we
discuss the yields of dileptons of the system. We give the results
and discussions in section IV. Finally, in Sec. V, we give brief
summary and conclusion.

\section{\label{sec:level2}EVOLUTION OF THE DISSIPATIVE QGP SYSTEM}
In this work, we describe the distribution  functions of partons
with J\"{u}ttner distributions
$f_{q(\bar{q})}$=${\lambda_{q(\bar{q})}}/{(e^{(p{\mp}\mu_{q})/T}+\lambda_{q(\bar{q})})}$
for quarks (antiquarks) and
$f_g$=${\lambda_g}/{(e^{p/T}-\lambda_g)}$ for gluons, where
fugacity $\lambda_i$ ($\leq$1) of the parton of type $i$ is
used to characterize the non-equilibrium of the system. Based on
these distribution functions, we first derive the thermodynamic
relations of the chemically equilibrating QGP system at finite
baryon density. Expanding densities of quarks (antiquarks)
\begin{eqnarray}
  n_{q(\bar q)}=\frac{g_{q(\bar q)}}{2\pi^2}\lambda_{q(\bar q)}\int\frac{p^2dp}{\lambda_{q(\bar q)}+e^(p\mp\mu_q)/T}
\end{eqnarray}
over quark chemical potential $\mu_q$, we get the baryon density
of the system [18]
\begin{eqnarray}&&
n_{b,q}=\frac{g_q}{6\pi^2}[T^3(Q_1^2\lambda_q-\bar{Q}_1^2\lambda_{\bar{q}})
+2\mu_qT^2(Q_1^1\lambda_q+\bar{Q}_1^1\lambda_{\bar{q}})\nonumber\\&&
+T\mu_q^2(Q_1^0\lambda_q-\bar{Q}_1^0\lambda_{\bar{q}})+\frac{1}{3}\mu_q^3
(\frac{\lambda_q}{\lambda_q+1}+\frac{\lambda_{\bar{q}}}{\lambda_{\bar{q}}+1}) ],
\end{eqnarray}
and the corresponding energy density
\begin{eqnarray}&&
\epsilon_{QGP} = \frac{g_q}{2\pi^2}
[T^4(Q_1^3\lambda_q+\bar{Q}_1^3\lambda_{\bar{q}})
+3\mu_qT^3(Q_1^2\lambda_q-\bar{Q}_1^2\lambda_{\bar{q}}) \nonumber \\ & &
+3\mu_q^2T^2(Q_1^1\lambda_q+\bar{Q}_1^1\lambda_{\bar{q}})
+T\mu_q^3(Q_1^0\lambda_q-\bar{Q}_1^0\lambda_{\bar{q}}) \nonumber \\&&
+\frac{1}{4}\mu_q^4(\frac{\lambda_q}{\lambda_q+1}+
\frac{\lambda_{\bar{q}}}{\lambda_{\bar{q}}+1})+\frac{g_g}{g_q}T^4G_1^3\lambda_g
+\frac{2\pi^2B_0}{g_q}],
\end{eqnarray}
where $g_{q(\bar{q})}$ and $g_g$ are degeneracy factors of quarks
(antiquarks) and gluons respectively. Since the convergence of the
following integral factors
 appearing in the expansion above
\begin{eqnarray} &&
G_m^n=\int\frac{Z^ndZ}{(e^Z-\lambda_g)^m}\quad\mbox{,}\quad
Q_m^n=\int\frac{Z^ndZ}{(e^Z+\lambda_q)^m} \nonumber\\ &&
\bar{Q}_m^n=\int\frac{Z^ndZ}{(e^Z+\lambda_{\bar{q}})^m}
\end{eqnarray}
is very rapid, it is easy to calculate these integral numerically
[18].

We consider the reactions leading to chemical equilibrium:
 $gg$$\rightleftharpoons$$ggg$ and
$gg$$\rightleftharpoons$$q\bar{q}$. Assuming that elastic parton
scatterings are sufficiently rapid to maintain local thermal
equilibrium, the evolutions of gluon and quark densities can be
given by the master equations, respectively. We first extend the
master equations to include the viscosity as done in [9].
Similarly, the evolution of baryon density can be described by a
corrected conservation equation of baryon number including a
viscous term. In addition, due to viscosity, a viscous term would
be contained in the conservation equation of the energy-momentum,
too. Combining the master equations together with the equation of
baryon number conservation and equation of energy-momentum
conservation including viscous corrections, for longitudinal
scaling expansion of the system, one can get a set of coupled
relaxation equations (CRE) describing evolutions of the
temperature $T$, quark chemical potential $\mu_q$, and fugacities
$\lambda_q$ for quarks and $\lambda_g$ for gluons on the basis of
the thermodynamic relations of the chemically equilibrating QGP
system at finite baryon density [9,18]
\begin{eqnarray} &&
(\frac{1}{{\lambda}_{g}}+\frac{G_{2}^{2}}{G_{1}^{2}})
{\dot{{\lambda}_{g}}}
+3\frac{\dot{T}}{T}+\frac{1}{{\tau}}
=R_3[1-\frac{G_1^2}{2\xi(3)}{\lambda}_g]   \nonumber\\ &&  -2R_2[1-{(\frac{2{\xi}(3)}{G_1^2})}^2
\frac{n_{g}n_{\bar{q}}}{\bar{n_g}\bar{n_{\bar{q}}}}\frac{1}{\lambda_g^2}]+\frac{\eta}{\epsilon{\tau}^2}
\end{eqnarray}

\begin{eqnarray} &&
\dot{\lambda_q}[T^3(Q_1^2-\lambda_{q}Q_2^2)+2\mu_{q}T^2(Q_1^1-\lambda_{q}Q_2^1)   \nonumber\\ &&
+T{\mu}_q^2(Q_1^0-\lambda_{q}Q_2^0)+\frac{1}{3}{\mu}_q^3\frac{1}{(\lambda_{q}+1)^2}]   \nonumber\\ &&
+\dot{T}[3\lambda_{q}T^2Q_1^2+4\lambda_{q}\mu_{q}TQ_1^1+\lambda_q\mu_q^2Q_1^0]  \nonumber\\ &&
+\dot{\mu_q}[2\lambda_{q}T^2Q_1^1+2\lambda_q\mu_qTQ_1^0+\mu_q^2\frac{\lambda_q}{\lambda_q+1}]
+\frac{n_q^0}{\tau}  \nonumber\\ &&
=n_g^0R_2[1-(\frac{2\xi(3)}{G_1^2})^2\frac{1}{\lambda_g^2}
\frac{n_{g}n_{\bar{q}}}{\bar{n_g}\bar{n_{\bar{q}}}}\frac{1}{\lambda_g^2}]
+\frac{\eta{n_q^0}}{\epsilon\tau^2}
\end{eqnarray}

\begin{eqnarray} &&
\dot{\lambda_q}[2T^2\mu_q(Q_1^1-\lambda_{q}Q_2^1)
+\frac{1}{3}{\mu}_q^3\frac{1}{(\lambda_{q}+1)^2}] \nonumber\\ &&
+\dot{T}4\lambda_{q}\mu_{q}TQ_1^1
+\dot{\mu_q}[2\lambda_{q}T^2Q_1^1+\mu_q^2\frac{\lambda_q}{\lambda_q+1}] \nonumber\\ &&
=-\frac{1}{\tau}[2T^2\mu_qQ_1^1\lambda_q+\frac{1}{3}\mu_q^3\frac{\lambda_q}{\lambda_q+1}]
+\frac{6{\pi}^2}{g_q}\frac{\eta{n_b}}{\epsilon\tau^2}
\end{eqnarray}

\begin{eqnarray} &&
\dot{\lambda_g}\frac{g_g}{g_q}T^4(G_1^3+\lambda_gG_2^3)
+\dot{\lambda_q}[2T^4(Q_1^3-\lambda_qQ_2^3) \nonumber\\ &&
+6T^2\mu_q^2(Q_1^1-\lambda_{q}Q_2^1)
+\frac{1}{2}{\mu}_q^4\frac{1}{(\lambda_{q}+1)^2}] \nonumber\\ &&
+\dot{T}[8\lambda_{q}T^3Q_1^3+12\lambda_{q}\mu_{q}^2TQ_1^1+4\frac{g_g}{g_q}
\lambda_gT^3G_1^3]\nonumber\\ &&
+\dot{\mu_q}[12\mu_q\lambda_{q}T^2Q_1^1+2\mu_q^3\frac{\lambda_q}{\lambda_q+1}]\nonumber\\ &&
=-\frac{4}{3\tau}[2T^4Q_1^3\lambda_q+6T^2\mu_q^2\lambda_qQ_1^1
+\frac{\mu_q^4}{2}\frac{\lambda_q}{\lambda_q+1}\nonumber\\ &&
+\frac{g_g}{g_q}\lambda_qT^4G_1^3]
+\frac{4}{3}\frac{2{\pi}^2}{g_q}\frac{\eta}{\tau^2}
\end{eqnarray}
where $\bar{n}_{q(\bar{q})}$ is the value of $n_{q(\bar{q})}$ at
$\lambda_{q(\bar{q})}=1$, $n_q^0=n_q/(g_q/2\pi^2)$,
$n_g^0=n_g/(g_g/2\pi^2)$, $\xi(3)$=1.20206, and $\eta$ the shear
viscous coefficient. The gluon and quark production rates
$R_3/T$ and $R_2/T$ are respectively given by[6,18-20]

\begin{eqnarray}
  R_3/T=\frac{32}{3a_1}\frac{\alpha_s}{\lambda_g}[\frac{(M_D^2+s/4)M_D^2}{9g^2T^4/2}]^2I(\lambda_g,\lambda_q,T,\mu_q),
\end{eqnarray}

\begin{eqnarray}
  R_2/T=\frac{g_g}{24\pi}\frac{G_1^{12}}{G_1^2}N_f\alpha_s^2\lambda_g\ln(\frac{1.65}{\alpha_s\lambda_g}),
\end{eqnarray}

\begin{eqnarray}
  M_D^2=\frac{3g^2T^2}{\pi^2}[2G_1^1\lambda_g+2N_fQ_1^1\lambda_q+N_f(\frac{\mu_q}{T})^2
  (\frac{\lambda_q}{\lambda_q+1})], \nonumber \\
\end{eqnarray}
where $M_D^2$ is the Debye screening mass, $g^2=4\pi\alpha_s$, and $I (\lambda_g,\lambda_q,T,\mu_q)$ is the
function of $\lambda_g$, $\lambda_q$, $T$, $\mu_q$, as used in
[5-6]. We here take the quark flavor $N_f$=2.5 [18-20]. Solving
the set of evolution equations (5)$\relbar$(8) under given initial
values obtained from Hijing model, we can get the evolutions of
temperature $T$, quark chemical potential $\mu_{q}$ and fugacities
$\lambda_{q}$ for quarks and $\lambda_{g}$ for gluons.

To discuss the effects of shear viscous coefficient, we have quoted two different expressions of it:  $\eta_1$ and $\eta_2$ are taken from [9,13], respectively.
\begin{eqnarray}
\eta_1=\eta_0\frac{\epsilon_{QGP}}{T}
\end{eqnarray}
\begin{eqnarray}
\eta_2=\frac{T}{\sigma_{\eta}}[\frac{n_g}{\frac{9}{4}n_g+n_q}
+\frac{n_q}{\frac{4}{9}n_q+n_g}]
\end{eqnarray}
where $\eta_0$ is treated as a constant [9], and $\sigma_{\eta}$
the transport cross section [13].

Now, we discuss the calculation of the viscous coefficient $\eta$ in our work. According to [13,14,21], the shear viscous
coefficient using the relativistic kinetic theory for a massless
QGP in the relaxation time approximation is written as:
\begin{equation}
\eta_i=\frac{4}{15}\epsilon_i\lambda_i,
\end{equation}
where $\lambda_i$ is the mean free path of particle of type $i$ in
QGP, which in a chemically equilibrating QGP for gluon is given by
\begin{equation}
\lambda_g=\frac{4}{9n_g}\frac{1}{2\pi{\alpha_s}^2}
\frac{M_D^2(M_D^2+9T^2/2)}{9T^2/2},
\end{equation}
and for quark by
\begin{equation}
\lambda_q=\frac{9}{4n_q}\frac{1}{2\pi{\alpha_s}^2}
\frac{M_D^2(M_D^2+9T^2/2)}{9T^2/2},
\end{equation}
where $M_D^2$ is the Debye screening mass and given by (11). Thus,
we can directly calculate the viscous coefficients $\eta_g$,
$\eta_g$ and their total $\eta$ from the thermodynamic quantities
of the QGP system.

\section{\label{sec:level3} DILEPTON PRODUCTION}

 Based on the evolution of the QGP system, we first consider dilepton production from
 quark annihilation $q\bar{q}{\rightarrow}gl\bar{l}$
and Compton scatterings $qg{\rightarrow}{q}l\bar{l}$ and
$\bar{q}g{\rightarrow}{\bar{q}}l\bar{l}$. Their production rates
can be calculated by [16]
\begin{eqnarray}
E{{dR}\over{d^{3}p}}&=& {{1}\over{2(2\pi)^{8}}}\int{{{d^{3}p_{1}}\over{2E_{1}}}
{{d^{3}p_{2}}\over{2E_{2}}}{{d^{3}p_{3}}\over{2E_{3}}}
 f_{1}(E_{1})f_{2}(E_{2})} (1\pm \nonumber\\
 &&  f_{3}(E_{3}))\cdot\delta^{4}(P_{1}+P_{2}-P_{3}
-K){\sum{\mid{M}\mid}^{2}},\nonumber\\
\end{eqnarray}
where $f(E)$ is the J\"uttner distribution function of partons,
$\sum{\mid{M}\mid}^{2}$ the square of the matrix element for
reaction processes summed over spins, colors and flavors. The plus
sign is for the annihilation process and the minus for the two
Compton processes. According to [22], above equation can be
rewriten as

\begin{eqnarray}
{{dR}\over{d^{2}M}}&=&{{100}\over{27}}{{\alpha^{2}\alpha_{s}}\over{\pi^{5}M}}
\int{dsdt{{u^{2}+t^{2}+2sM^{2}}\over{ut}}}\lambda^{2}_{q}  \nonumber\\
&&
\int{{dE_{1}}\over{e^{(E_{1}-\mu_{q})/T}+\lambda_{q}}}{{dE_{2}}
\over{e^{(E_{1}+\mu_{q})/T}+\lambda_{q}}}
\int{{dE}\over{E}}[1+   \nonumber\\
&& {{\lambda_{g}}\over{e^{(E_{1}+E_{2}-E)/T}-\lambda_{g}}}]
{{\theta(P(E_{1},E_{2}))}\over{(P(E_{1},E_{2}))^{1/2}}}
\end{eqnarray}

for the annihilation process, and
\begin{eqnarray}
{{dR}\over{d^{2}M}}&=&{20}{{\alpha^{2}\alpha_{s}}\over{\pi^{5}M}}
\int{dsdt{{u^{2}+s^{2}+2tM^{2}}\over{-us}}}\lambda_{q}\lambda_{g} \nonumber\\
&&
\int{{dE_{1}}\over{e^{(E_{1}\mp\mu_{q})/T}+\lambda_{q}}}{{dE_{2}}\over{e^{(E_{1})/T}-\lambda_{g}}}
\int{{dE}\over{E}}[1-   \nonumber\\
&&
{{\lambda_{q}}\over{e^{(E_{1}+E_{2}-E\mp\mu_{q})/T}+\lambda_{q}}}]
{{\theta(P(E_{1},E_{2}))}\over{(P(E_{1},E_{2}))^{1/2}}}
\end{eqnarray}
for Compton scatterings, where
$P(E_{1},E_{2})=-(tE_{1}+(s+t)E_{2})^{2}+2Es((s+t)E_{2}-tE_{1})-s^{2}E^{2}+s^{2
}t+st^{2}$, $\theta$ is the step function, $\alpha$ the
fine-structure constant, and $\alpha_{s}$ the running coupling
constant. The minus sign is for the Compton process
$qg{\rightarrow}{q}l\bar{l}$ and the plus for
$\bar{q}g{\rightarrow}{\bar{q}}l\bar{l}$. The letters $s$, $t$ and
$u$ are the Mandelstam variables. And the integration are
performed over $-(s-M^2)+k^{2}_{c}\leq{t}\leq-k^{2}_{c}$ and
$M^2+2k^{2}_{c}\leq{s}< {\infty}$ [16]. The cutoff $k^{2}_{c}$ is
replaced by the thermal quark mass 2$m^{2}_{q}$ [22]. For a
chemically equilibrating QGP system at finite baryon density
$m_{q}^{2}$ is given by
\begin{eqnarray}
m_{q}^{2}=\frac{4\alpha_{s}}{3\pi}T^{2}[2(G_{1}^{1}\lambda_{g}+Q_{1}^{1}\lambda_{q})
+(\frac{\mu_{q}}{T})^{2}\frac{\lambda_{q}}{\lambda_{q}+1}]
\end{eqnarray}
where integral factors $G_{1}^{1}$ and $Q_{1}^{1}$ have been given above.

Obviously, above calculations give up the infrared contribution because of introducing the infrared cutoff $k_c^2$.
Authors of [22] have discussed the infrared contribution to photon production.
Following their calculation, in this work, we have given an assessment of the
contribution from the infrared part to the dilepton produnction.
The dilepton production rate with total energy $E$ and total momentum $\mathbf{p}$ can be calculated by [22,23]
\begin{eqnarray}
  \frac{dR}{dEd^3p}= \frac{\alpha}{12\pi^4}\frac{1}{P^2}\frac{1}{e^{E/T}-1}\mathrm{Im}
  \Pi^{\ \ \mu}_{R,\mu},
\end{eqnarray}
where $\Pi^{\mu\nu}_{R}$ is the retarded photon self-energy.
The infrared divergence mentioned above is caused by propagation of soft momenta. To cure the problem it is necessary to dress one of the quark propagators (as done in Fig. 3 of [22]).
Then the retarded photon self-energy can be written as [22]
\begin{eqnarray}&&
\Pi^{\mu\nu}(p)=\nonumber \\&&
-\frac{5}{3}e^2T\sum_{k_0}\int\frac{d^3k}{(2\pi)^3}Tr[S^{\ast}(k)\gamma^{\mu}S(p-k)\gamma^{\nu}],
\end{eqnarray}
where $S^{\ast}(k)$ is the dressed propagator for a quark with four-momentum $k$, and $S(q)$ is the bara propagator for a quark with four-momentum $q=p-k$.
Following [22], after performing some calculations and applying the elegant method developed by Braaten, Pisarski and Yuan [23], one can finally obtain a simplified expression of the imaginary part of the retarded photon self-energy
\begin{eqnarray}&&
\mathrm{Im}\Pi^{\ \ \mu}_{R,\mu}=
\frac{5e^2}{6\pi}(e^{E/T}-1)\nonumber \\&&
\int_0^{k_c}dk\int_{-k}^kd\omega f(\omega)f(E-\omega)(k-\omega)\beta_{+}(\omega,k),
\end{eqnarray}
where $f$ is the J\"uttner distribution function of partons,
and
\begin{eqnarray}&&
  \beta_{+}(\omega,k)= \nonumber \\&&
  \frac{\frac{1}{2}m_q^2(k-\omega)}
  {\{k(\omega-k)-m_q^2[Q_0(z)-Q_1(z)]\}^2+[\frac{1}{2}\pi m_q^2(1-z)]^2},\nonumber\\
\end{eqnarray}
where $z=\omega/k$ [24], $Q_0$ and $Q_1$ are the Legendre functions of the second kind.
After performing integral over the dilepton energy ($E\ge M$) one can obtain
the dilepton production rate over square invariant mass of dileptons $dR/dM^2$.

Similar to the preceding treatment, the production rate of
quark-antiquark annihilation $q\bar{q}{\rightarrow}l\bar{l}$ can
be given by [25]
\begin{eqnarray}
\frac{dR}{dM^2}&=&\frac{5}{24\pi^4}M^2\sigma_{l\bar l}(M^2)\nonumber\\ &&
\times\int_0^\infty dp_1f_q(p_1)
\int_{M^2/4p_1}^\infty dp_2f_q(p_2),
\end{eqnarray}
where $\sigma_{l\bar l}(M^2)=\frac{20}{3}4\pi\alpha^2/3M^2$ is the
quark annihilation cross section.

Aurenche, Gelis, and their co-workers have studied the dilepton
production from bremsstrahlung and off-shell annihilation where the quark undergoes multiple scattering in the medium as shown
in Fig.1 of [26]. Their calculation includes the Landau-Pomeranchuk-Migdal effect and is concluded that this contribution is
important. According to these authors' approach,
the imaginary part of the retarded current-current correlator
$\Pi_{R,\mu}^{\ \ \mu}(P)$ in Eq.(21) can be computed by
\begin{eqnarray} & &
  \mathrm{Im}\Pi_{R,\mu}^{\ \ \mu}\approx\frac{5}{6\pi}\int_{-\infty}
  ^{+\infty}dq_0[f(k_0)-f(q_0)]\times\mathrm{Re}\int\frac{d^2\bm{q}_{\perp}}{(2\pi)^2} \nonumber \\ & &
  \left[\frac{q_0^2+k_0^2}{2(q_0k_0)^2}\bm{q}_{\perp}\cdot\bm{f}(\bm{q}_{\perp})
  +\frac{1}{\sqrt{|q_0k_0|}}\frac{P^2}{p^2}g(\bm{q}_{\perp})\right],
\end{eqnarray}
with $k_0\equiv q_0+E$, $f$ is the J\"uttner distribution
function of partons again, and the dimensionless functions
$\bm{f}(\bm{q}_{\perp})$ and $g(\bm{q}_{\perp})$ respectively obey the integral equations [26]
\begin{eqnarray}
  2\bm{q}_{\perp}=i\delta E\bm{f}(\bm{q}_{\perp})+\frac{4}{3}g_s^2T
  \int\frac{d^2\bm{l}_{\perp}}{(2\pi)^2}\mathcal{C}(\bm{l}_{\perp}) \nonumber \\
  \times[\bm{f}(\bm{q}_{\perp})-\bm{f}(\bm{q}_{\perp}+\bm{l}_{\perp})]
\end{eqnarray}
and
\begin{eqnarray}
  2\sqrt{|q_0k_0|}=i\delta Eg(\bm{q}_{\perp})+g_s^2C_FT\int\frac{d^2\bm{l}_{\perp}}{(2\pi)^2}
  \mathcal{C}(\bm{l}_{\perp})\nonumber \\
  \times[g(\bm{q}_{\perp})-g(\bm{q}_{\perp}+\bm{l}_{\perp})].
\end{eqnarray}
Using the method described in [26], we recast (27) and (28) as differential equations and solve them using a simple algorithm,
finally, can get
$Re{\int}d^2\bm{q}_{\perp}/(2\pi)^2\bm{q}_{\perp}\cdot\bm{f}(\bm{q}_{\perp})$
 and $Re{\int}d^2\bm{q}_{\perp}/(2\pi)^2g(\bm{q}_{\perp})$ and the corresponding dilepton production rate of multiple scattering process.

For the QGP system, produced in collisions at RHIC energies, with
very high initial temperature ($\approx$0.57 GeV) [6,18], thermal
charmed quark production and its contribution to lepton pairs
should be contained, especially, those from the gluon fusion
$gg{\rightarrow}c\bar{c}$ and quark-antiquark annihilation
$q\bar{q}{\rightarrow}c\bar{c}$.
 Similar to the calculation
for $q\bar{q}{\rightarrow}l\bar{l}$, replacing the cross section
${\sigma}_{l\bar l}(M^2)$ appears in the expression (25) with those
of the reactions $q\bar{q}{\to}c\bar{c}$ and $gg{\to}c\bar{c}$ in
leading order QCD, we can compute the yields of charm pairs in the
QGP. Almost all of the produced thermal charmed quarks would
eventually hadronize to $D$-mesons [17]. Considering that the
$D$-meson decays to leptons with a 17$\%$ branching ratio for
charged $D$-mesons [17,27,28], finally one can obtain the
contribution of charmed quarks from reactions
$gg{\rightarrow}c\bar{c}$ and $q\bar{q}{\rightarrow}c\bar{c}$ to
lepton pairs.

We integrate these production rates over the space-time
volume of the reaction. According to Bjorken's model, the volume element is $d^4x=d^2x_Tdy \tau d\tau$,
where $\tau$ is the evolution time of the system and $y$ the rapidity of the fluid element.
We consider $Au^{197}+Au^{197}$ central collisions, so the integration
over transverse coordinates just yields a factor of
$d^2x_T=\pi R_A^2$, where $R_A$ is the nuclear radius. Finally we
obtain the dilepton spectra of the system
\begin{equation}
\frac{dN}{dydM^2}=\pi R^2_A\int \tau d\tau\frac{dR}{dM^2}.
\end{equation}

\section{\label{sec:level4}CALCULATED RESULTS AND DISCUSSION}

\begin{figure}
\vspace{-0.1truein}
\includegraphics[width=8.4cm]{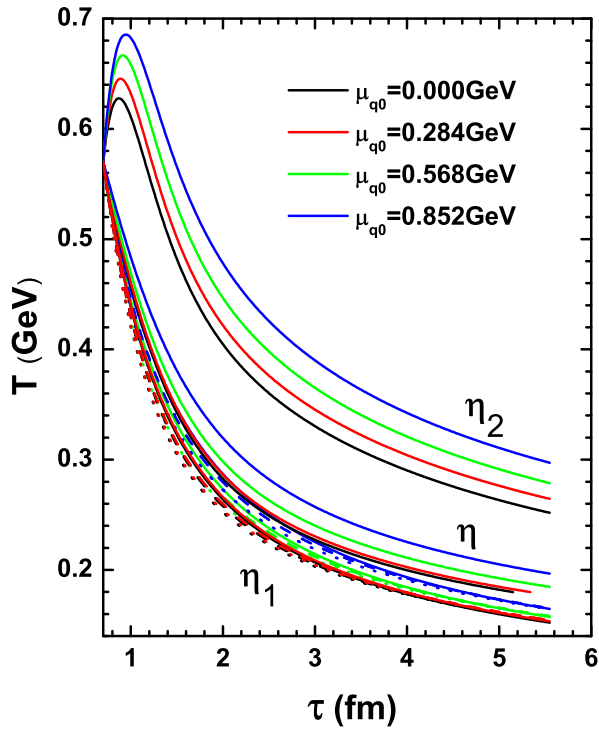}
\vspace{-0.1truein} \caption{\footnotesize (Color online) The evolutions of the
temperature $T$ with proper time $\tau$ for viscous coefficients
$\eta_1$, $\eta_2$, and $\eta$ at initial quark chemical
potentials $\mu_{q0}$=0.000, 0.284, 0.568, and 0.852 GeV. The
$\eta_1$ is the viscous coefficients for parameters $\eta_0$=0.0,
0.4, and 0.8 as given in [9], and the corresponding values are
denoted by the solid, dash and dot curves, respectively.
}\label{fig1}
\end{figure}

In this work, we focus on discussing $Au^{197}+Au^{197}$ central
collisions at the RHIC energies. In order to compare with
[9,29,30], we take initial values of the system: $\tau_0$=0.70 fm,
$T_0$=0.570 GeV, $\lambda_{g0}$=0.08 and $\lambda_{q0}$=0.02 from
the Hijing model calculation. We have investigated the effect of
viscosity on the evolution of the system using those expressions
of the shear viscosity as shown in (12)$\relbar$(16). To
understand the effect of the baryon density on the dilepton
production, we have solved the CRE for initial quark chemical
potentials $\mu_{q0}$=0.000, 0.284, 0.568, and 0.852 GeV for given
viscous coefficients,
 and obtained the evolutions of the temperature $T$, quark chemical potential
$\mu_{q}$ and fugacities $\lambda_{g}$ and $\lambda_{q}$ of the
system. In Fig.1, we have shown the evolutions of the temperature
$T$ with proper time $\tau$ for the various viscous cofficients
from the expressions in [9,13]. Following [9], we take the
parameters $\eta_0$=0.0, 0.4, and 0.8. The corresponding curves
are indicated by the solid, dash and dot lines, respectively. We
can see that the viscosity leads to the increase of the
temperature of the QGP system. We also note that the temperature
is unreasonably heightened for viscous coefficient $\eta_2$. For $\eta_1$, the evolution of temperature seems to be
reasonable, however, the viscosity $\eta_1$ is only obtained
through adjusting the parameter $\eta_0$. In this work, we have
calculated the viscous coefficients $\eta$ by the thermodynamic
quantities of the QGP system using (14)$\relbar$(16). From Fig.1,
we note that the calculated temperature distribution is
reasonable. In addition, the temperature is also a increasing
function of the initial quark chemical potential. Fig.2 shows the
value of $\eta$ as a function of the initial temperature, where
the black, red, green and blue curves denote, in turn, the
calculated $\eta$ for initial quark chemical potentials
$\mu_{q0}$=0.000, 0.284, 0.568, and 0.852 GeV at initial values
mentioned above. From Fig.2, one can see that $\eta$ increases
with increasing the temperature $T$.

\begin{figure}
\vspace{-0.1truein}
\includegraphics[width=8.4cm]{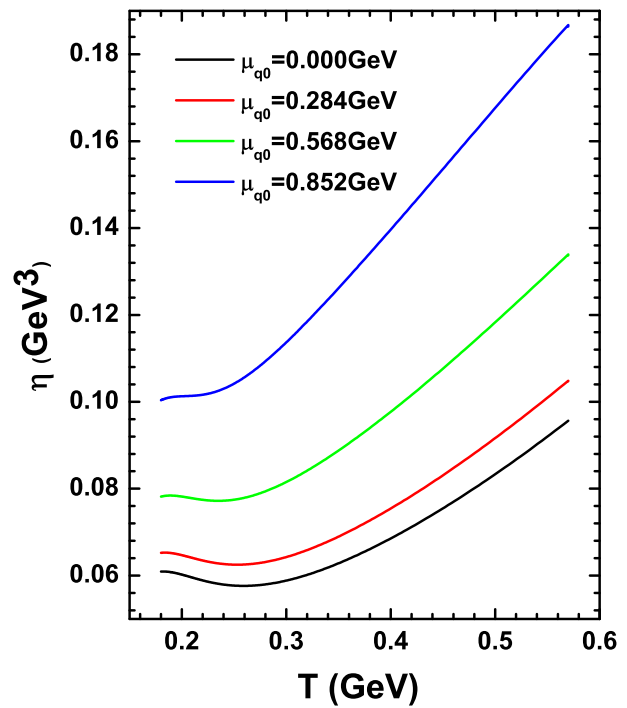}
\vspace{-0.1truein} \caption{\footnotesize (Color online) The viscous coefficient
$\eta$ as a function of the temperature. Black, red, green and
black denote, in turn, the calculated viscous coefficients $\eta$
for initial quark chemical potentials $\mu_{q0}$=0.000, 0.284,
0.568, and 0.852 GeV at the same initial conditions as given in
Fig.1. }\label{fig2}
\end{figure}

\begin{figure}
\vspace{-0.1truein}
\includegraphics[width=8.4cm]{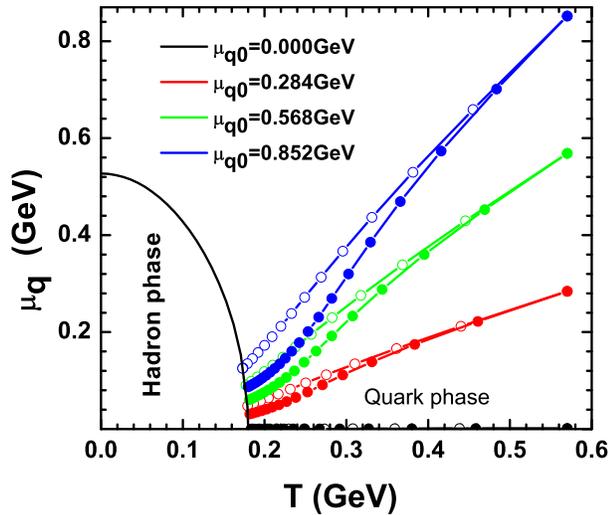}
\vspace{-0.1truein} \caption{\footnotesize (Color online) The calculated
evolution paths of the system in the phase diagram for the initial
values as given in Fig.1, where black, red, green and blue curves
are, in turn, the calculated evolution paths for initial quark
chemical potentials $\mu_{q0}$=0.000, 0.284, 0.568, and 0.852 GeV.
The lines with open circles denote the evolution of system without
viscous effect while the lines with solid circles are the
evolution of the system with viscosity. The time interval between
the two circles is 0.3 fm. }\label{fig3}
\end{figure}

\begin{figure}
\vspace{-0.1truein}
\includegraphics[width=8.4cm]{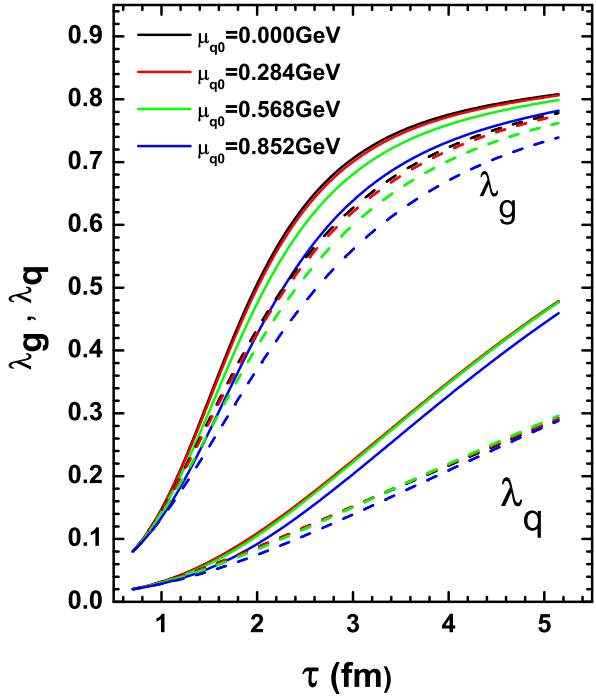}
\vspace{-0.1truein} \caption{\footnotesize (Color online) The calculated
equilibration rates at the same initial conditions as given in
Fig.1. The solid lines are for the cases of viscosity while short
dash lines for the ideal cases, where the black, red, green and
blue lines are the calculated values for initial quark chemical
potentials $\mu_{q0}$=0.000, 0.284, 0.568, and 0.852 GeV. }\label{fig4}
\end{figure}

The estimated evolution paths of the system in the phase diagram
have been shown in Fig.3, where black, red, green and blue curves
are, in turn, the calculated evolution paths for initial quark
chemical potentials $\mu_{q0}$=0.000, 0.284, 0.568, and 0.852 GeV.
 The solid line is the phase boundary
between the quark phase and hadronic phase. The lines with open
circles denote the evolution of system without viscous effect,
while the lines with solid circles are from the system with
viscosity. The time interval between the two circles is 0.3 fm.
The corresponding equilibration rates of gluons and quarks,
$\lambda_{g}$ and $\lambda_{q}$, are shown in Fig. 4. The solid
lines are for the cases of viscosity, and the short dash lines
denote ideal cases, where the black, red, green and blue curves
are, in turn, the calculated values for initial quark chemical
potentials $\mu_{q0}$=0.000, 0.284, 0.568, and 0.852 GeV. From
Figs.3 and 4 we see that the evolution of the system becomes
slower due to viscosity, whereas the equilibration rate of the
plasma becomes faster compared to the one in the ideal case. And
also the effect of the initial quark chemical potential on the
evolution is in accordance with previous conclusion [18].

\begin{figure}
\vspace{-0.1truein}
\includegraphics[width=8.4cm]{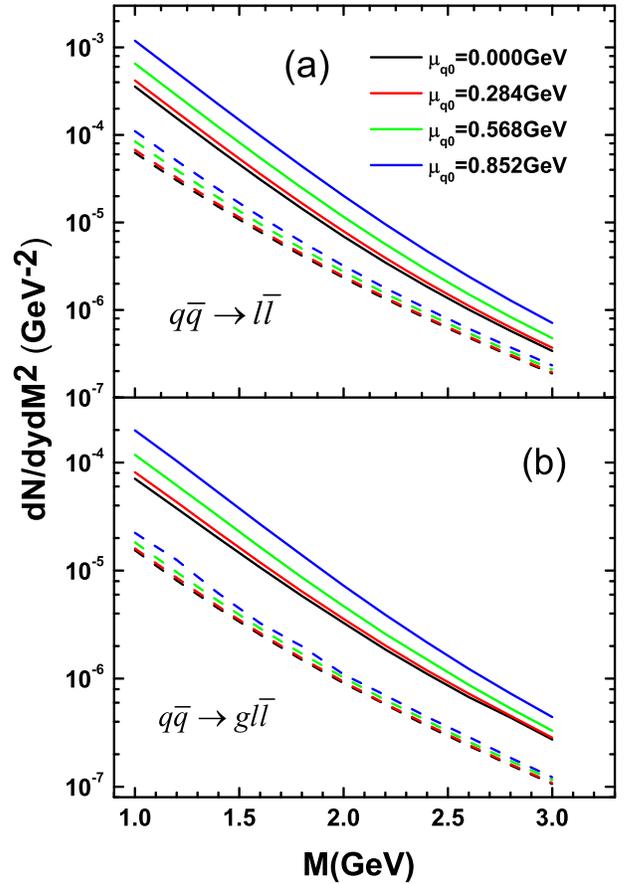}
\vspace{-0.1truein} \caption{\footnotesize (Color online) The calculated
dilepton spectra from quark annihilation processes $q\bar{q}$$\rightarrow$$l\bar{l}$
 and $q\bar{q}$$\rightarrow$$gl\bar{l}$. The signs of the curves are the same
 as those in Fig.4.}\label{fig5}
\end{figure}

\begin{figure}
\vspace{-0.1truein}
\includegraphics[width=8.4cm]{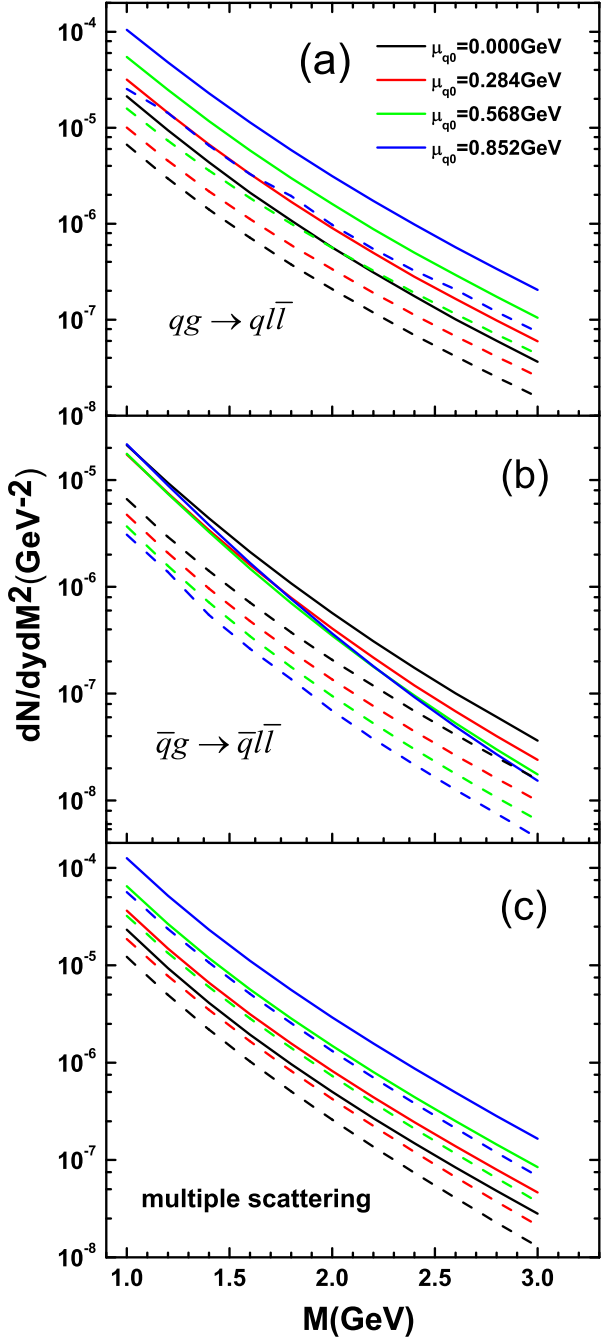}
\vspace{-0.1truein} \caption{\footnotesize (Color online) The calculated
dilepton spectra from processes $qg$$\rightarrow$$ql\bar{l}$,
  $\bar{q}g$$\rightarrow$$\bar{q}l\bar{l}$ and multiple scattering. The signs of the curves are the same
 as those in Figs. 4 and 5.}\label{fig6}
\end{figure}

\begin{figure}
  \vspace{-0.1truein}
\includegraphics[width=8.4cm]{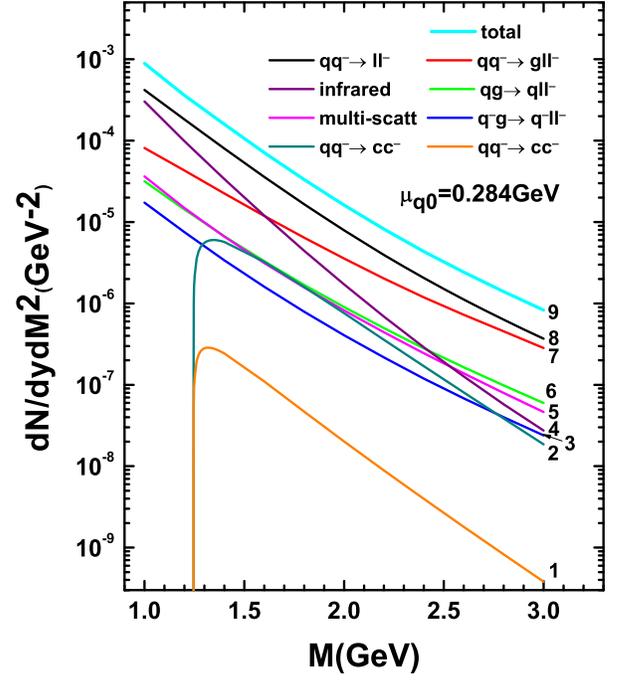}
\vspace{-0.1truein} \caption{\footnotesize (Color online) The calculated spectra
of all processes for the initial quark chemical potential
$\mu_{q0}$=0.284 GeV. Curves 1$\relbar$9 are, in turn, the
calculated spectra for $q\bar{q}$$\rightarrow$$c\bar{c}$,
$gg$$\rightarrow$$c\bar{c}$,
$\bar{q}g$$\rightarrow$$\bar{q}l\bar{l}$, soft dileptons,
multiple scattering, $qg$$\rightarrow$$ql\bar{l}$,
$q\bar{q}$$\rightarrow$$gl\bar{l}$,
$q\bar{q}$$\rightarrow$$l\bar{l}$, and their total. }\label{fig7}
\end{figure}

\begin{figure}
\vspace{-0.1truein}
\includegraphics[width=8.4cm]{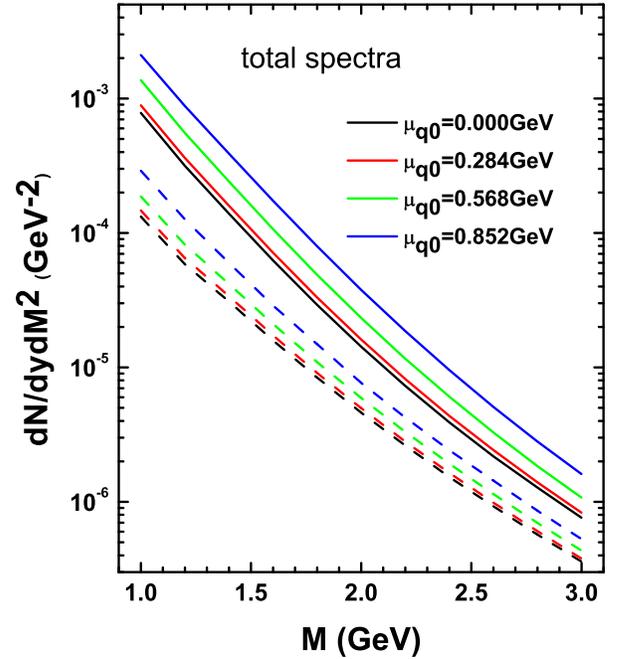}
\vspace{-0.1truein} \caption{\footnotesize (Color online) The calculated total
dilepton spectra of the system.
The signs of the curves are the same
 as those in above figures.}\label{fig8}
\end{figure}

Based on the evolution of the system described in Figs.3
 and 4, we have calculated dilepton
spectra for processes $q\bar{q}$$\rightarrow$$l\bar{l}$,
$q\bar{q}$$\rightarrow$$gl\bar{l}$, $qg$$\rightarrow$$ql\bar{l}$,
$\bar{q}g$$\rightarrow$$\bar{q}l\bar{l}$,
$q\bar{q}$$\rightarrow$$c\bar{c}$, $gg$$\rightarrow$$c\bar{c}$,
and multiple scattering at the four initial quark chemical
potentials as given above. The calculated dilepton spectra from
quark annihilation processes $q\bar{q}$$\rightarrow$$l\bar{l}$
 and $q\bar{q}$$\rightarrow$$gl\bar{l}$ are, in turn,
 shown in panels (a) and (b) of Fig.5. The solid lines are the
 spectra of the system with the effect of viscosity while dash lines indicate those from ideal system,
where the black, red, green and blue lines denote the calculated
spectra for initial quark chemical potentials $\mu_{q0}$=0.000,
0.284, 0.568, and 0.852 GeV.
 One can see that the calculated spectrum goes up with increasing
the initial quark chemical potential. The law is valid for
processes $qg$$\rightarrow$$ql\bar{l}$ and multiple scattering as
shown in panels (a) and (c) in Fig.6. We know that the quark
density goes up with the increase of quark chemical potential
while the density of antiquark goes down, which will lead to a
suppression of the production from process
$\bar{q}g$$\rightarrow$$\bar{q}l\bar{l}$, as shown in panel (b) of
Fig.6. And also we can see from Figs.5 and 6 that the dilepton
production is a increasing function of viscosity, which would be
mainly attribute to the increase of the quark phase life-time due
to the viscosity.

The productions of soft dileptons, which are connected with the
infrared contribution, have been computed following the method
represented in [23] for initial quark chemical potentials
$\mu_{q0}$=0.000, 0.284, 0.568, and 0.852 GeV. In Fig.7 we have
shown the results from all processes and their total for the
initial quark chemical potential $\mu_{q0}$=0.284 GeV. Curves
1$\relbar$9 represent, in turn, the calculated spectra for
$q\bar{q}$$\rightarrow$$c\bar{c}$,
$gg$$\rightarrow$$c\bar{c}$,
$\bar{q}g$$\rightarrow$$\bar{q}l\bar{l}$, soft dileptons,
multiple scattering, $qg$$\rightarrow$$ql\bar{l}$,
$q\bar{q}$$\rightarrow$$gl\bar{l}$,
$q\bar{q}$$\rightarrow$$l\bar{l}$, and their total. From Fig.7, one
can see that the spectra from the quark-antiquark annihilations
$q\bar{q}$$\rightarrow$$l\bar{l}$ and
$q\bar{q}$$\rightarrow$$gl\bar{l}$ dominate. The infrared
contribution is as important as that of reaction
$q\bar{q}$$\rightarrow$$gl\bar{l}$ and even higher than the later one in the range of small invariant
mass. The contributions from Compton-like scattering
$qg$$\rightarrow$$ql\bar{l}$, multiple scattering, and annihilation
$gg$$\rightarrow$$c\bar{c}$ can not also be neglected.

We have also given the total yields of all processes of the system
for initial conditions mentioned above, as shown in Fig. 8. The
black, red, green and blue curves represent the total yields for
$\mu_{q0}$=0.000, 0.284, 0.568, and 0.852 GeV, respectively. To
understand the effect of viscosity on the dileptons production, we
also give the yields for ideal QGP system, which are denoted by
dash lines. It shows clearly that the dilepton yield of the system
goes up with increasing initial quark chemical potential. However,
previous authors have found that dileptons produced in a
thermodynamic equilibrium QGP system are suppressed with
increasing initial quark chemical potential [1]. In this work,
since that both the quark chemical potential and the temperature
of the system are functions of time, compared with the baryon-free
QGP it necessarily takes a long time for value ($\mu_{q}$,$T$) of
the system to reach a certain point of the phase boundary to make
the phase transition. Furthermore, in the calculation we have
found that with increasing the initial quark chemical potential
the production rate of gluons goes up, and thus their
equilibration rate goes down, leading to the little energy
consumption of the system, i.e., slow cooling of the system. Since
gluons are much more than quarks in the system, with increasing
the initial quark chemical potential the cooling of the system
further slows down. These cause the quark phase life-time to
further increase, as seen in Fig.3. These effects will heighten
the dilepton yield and compensate the dilepton suppression,
leading the spectrum of the system to be an increasing function of
the initial quark chemical potential.  On the other hand, as seen
in Fig.3, due to the viscosity the evolution of the system becomes
even slower, so that the dilepton yield will be heightened, as
seen in Fig.8. From Fig.8 one can note that the dilepton yields
are remarkably heightened due to the effect of the viscosity of
the QGP system.

\section{\label{sec:level5}SUMMARY and CONCLUSION}

In this work, taking into account reactions
$gg$$\rightleftharpoons$$ggg$ and
$gg$$\rightleftharpoons$$q\bar{q}$ leading to the chemical
equilibrium of the QGP system, and conservations of
energy-momentum and baryon number, as well as viscosity of the QGP
system, we have derived a set of coupled CRE of the chemically
equilibrating QGP system with viscosity at finite baryon density,
produced from $Au^{197}+Au^{197}$ central collisions at RHIC
energies, which describes the space-time evolution of the system.
Then, we have solved the CRE, and directly obtained the viscous
coefficients from the thermodynamic quantities of the QGP system.
We note that the calculated results of the viscous coefficients
are reasonable. Subsequently, based on the evolution of the QGP
system we have computed the dilepton spectra of the QGP system, we
have found that the spectra is dominated by the quark-antiquark
annihilation $q\bar{q}$$\rightarrow$$l\bar{l}$ and
$q\bar{q}$$\rightarrow$$gl\bar{l}$, following by the multiple
scattering of quarks, compton-like scattering
$qg$$\rightarrow$$ql\bar{l}$ and annihilation
$gg$$\rightarrow$$c\bar{c}$. While we have also calculated the
infrared contribution and found it very important at the range of
small invariant mass of dileptons. Furthermore, we note that the
increase of the dilepton yield with increasing the initial quark
chemical potential can compensate the dilepton suppression, thus
eventually leading to the dilepton spectrum to be an increasing
function of the initial quark chemical potential. Especially, we
have found that the dilepton yield of the system is obviously
enhanced due to the viscous effect because this effect makes the
evolution of the system slow down and thus the life-time of the
QGP system increases.

\section{ACKNOWLEDGEMENTS}
This work is supported in part by THE Knowledge Innovation Project of
Chinese Academy of Science under Grant No. KJCX2-N11, CAS master scholar fund, the National Natural Science Foundation of China under Grant Nos 10075071, 10605037 and 10875159, the Major State Basic Research Development
Program in China under Contract No. G200077400.

\bigskip
\bigskip
\par

\bigskip
\end{document}